\documentclass[12pt]{article}
\input{epsf.sty}

\def\ignore#1{{}}

\renewcommand{\baselinestretch}{1.25}

\newcommand{\beeq}{\begin{equation}}
\newcommand{\eneq}{\end{equation}}
\newcommand{\beqn}{\begin{eqnarray}}
\newcommand{\eeqn}{\end{eqnarray}}

\def\la{\raise.16ex\hbox{$\langle$}\lower.16ex\hbox{}  }
\def\ra{\, \raise.16ex\hbox{$\rangle$}\lower.16ex\hbox{} }
\def\go{\rightarrow}

\def\psibar{ \psi \kern-.65em\raise.6em\hbox{$-$} \lower.6em\hbox{} }
\def\psibaralpha{ \psi^{(\alpha)} \kern-1.9em\raise.6em\hbox{$-$}
\kern+1.2em\hbox{}}
\def\psibara{ \psi^{(a)} \kern-1.9em\raise.6em\hbox{$-$}\kern+1.2em\hbox{}}

\def\eff{{\rm eff}}

\def\ep{\epsilon}

\def\justbox#1{$\vcenter{\hrule\hbox{\vrule\kern0pt
     \vbox{\kern0pt\hbox{#1}\kern0pt}\kern0pt\vrule}\hrule}$}
\def\boxit#1{$\vcenter{\hrule\hbox{\vrule\kern3pt
     \vbox{\kern3pt\hbox{#1}\kern3pt}\kern3pt\vrule}\hrule}$}
\def\bigbox#1{$\vcenter{\hrule\hbox{\vrule\kern5pt
     \vbox{\kern5pt\hbox{#1}\kern5pt}\kern5pt\vrule}\hrule}$}
\def\Bigbox#1{$\vcenter{\hrule\hbox{\vrule\kern7pt
     \vbox{\kern7pt\hbox{#1}\kern7pt}\kern7pt\vrule}\hrule}$}
\def\hugebox#1{$\vcenter{\hrule\hbox{\vrule\kern9pt
     \vbox{\kern9pt\hbox{#1}\kern9pt}\kern9pt\vrule}\hrule}$}
\def\Hugebox#1{$\vcenter{\hrule\hbox{\vrule\kern20pt
     \vbox{\kern20pt\hbox{#1}\kern20pt}\kern20pt\vrule}\hrule}$}
\def\thinbox#1{$\vcenter{\hrule\hbox{\vrule\kern1pt
     \vbox{\kern1pt\hbox{#1}\kern1pt}\kern1pt\vrule}\hrule}$}
\def\verythinbox#1{$\vcenter{\hrule\hbox{\vrule\kern.2pt
     \vbox{\kern.2pt\hbox{#1}\kern.2pt}\kern.2pt\vrule}\hrule}$}

\begin{document}

\thispagestyle{empty}

\baselineskip=16pt plus 1pt minus 1pt

\vskip 3cm

\begin{center}

{\Large\bf Solitons in the false vacuum\footnote{\it To appear in
the Proceedings of the YITP workshop ``Fundamental Problems and
Applications of Quantum Field Theory'', Kyoto, December 20 - 22,
2000.}}\\

\vspace{.5cm}

{\large    Yutaka Hosotani}\\
\vspace{.1cm}
{\it \small Department of Physics, Osaka University, Toyonaka, Osaka
560-0043}\\   
\end{center}


\begin{abstract}
\small
\baselineskip=12pt
When a potential for a scalar field has  two local minima , there arises
structure of spherical shells due to  gravitational interactions. 
\end{abstract}

\vskip .2cm

\baselineskip=14pt plus 1pt minus 1pt

Gravitational interactions, inherently attractive for ordinary
matter, can produce soliton-like objects even when such things
are strictly forbidden in flat space.  They become possible as
a consequence of the balance between repulsive and attractive forces.
One such example is a monopole or dyon solution in the pure
Einstein-Yang-Mills theory in the asymptotically anti-de Sitter
space.\cite{Hosotani1}  In this talk I report a new  shell
structure in a simple real scalar field theory.
 
In a scalar field theory given by
\beeq
{\cal L} = 
{1\over 16\pi G} ~ R
   + {1\over 2}  \phi_{;\mu} \phi^{;\mu}   - V[\phi]
\eneq
we seek a static, spherically symmetric spacetime for which
the metric can be written as
\beeq
ds^2 = 
\frac{H(r)}{p^2(r)}dt^2 - \frac{dr^2}{H(r)} -r^2 d\Omega^2 ~.
\label{metric1}
\eneq
We suppose that the potential $V(\phi)$ has
two minima at $f_1$ and $f_2$ separated by a barrier.  Einstein equations
and matter equation reduce to 
\beqn
&&\hskip -.8cm
\phi''(r) + \Gamma_\eff (r)  \phi'(r) = {1\over H} ~ V'[\phi]
~~,~~ \Gamma_\eff= {2\over r} + 4\pi G r \phi'^2 + {H'\over H} ~~,
\label{matter1}  \\
&&\hskip -.8cm
H = 1 - {2GM\over r} 
~~,~~
M(r) =\int_0^r 4\pi r^2 dr \, \left\{ {1\over 2} H \phi'^2 
   +V[\phi] \right\} ~~.
\label{metric2}
\eeqn

\vskip .5cm

\noindent
\bigbox{I. ~False vacuum black hole}
\vskip .3cm

Suppose that $V(f_1) = \ep >0$ and $V(f_2) =0$. See figure 1(a).
  $\phi= f_1$ and 
$\phi= f_2$ correspond to the false and true vacuum, respectively.  If the
universe  is in the false vacuum, a bubble of the true vacuum is created
by quantum tunneling which expands with accelerated velocity.  The
configuration is called a bounce.\cite{Coleman}  The bounce is
a Minkowski bubble in de Sitter.

Let us flip the configuration.\cite{Hosotani2}  The universe is in the
true vacuum and the inside of a sphere is excited to the false vacuum.
Is such a de Sitter lump in Minkowski possible?  If the lump is too
small, then it would be totally unstable.  The energy localized inside
the lump can dissipate to space infinity.  If the lump is big enough,
the Schwarzschild radius becomes larger than the lump radius, i.e.\
the lump is inside a black hole.  The energy cannot escape to infinity.
It looks like a soliton in the Minkowski space.  However, as being
a black hole, it is a dynamic object.  The configuration is
essentially time-dependent.

The critical configuration of size $R$ carries interesting numbers. 
The transition region located around $R$ is approximated by a thin wall.
The inside is a de Sitter space with $H = 1 - (r/a)^2$, $a^2 = 3/8\pi
G \ep$.  The outside is described by the  Schwarzschild metric with $H
= 1-(r_S/r)$, $r_S = 2G M = R^3/a^2$.  At the critical radius
$R_c=r_S=a=(3/8\pi G \ep)^{1/2}$; the Schwarzschild radius and
cosmological horizon coincide.  If the energy scale, $\ep^{1/4}$, is
at the GUT scale ($10^{15}$GeV), then
$R_c \sim 10^{27}\,$m and $M_c \sim 1$ kg.    
If  $\ep^{1/4} \sim 1 \,$GeV, then $R_c \sim 1 \,$km and 
$M_c \sim 10^5 \cdot M_{\rm sun}$, a typical mass of a black hole 
located near the center of each galaxy.  
If  $\ep^{1/4} = 2.4\,$meV as suggested from the estimated value of  the
present cosmological constant, then $R_c \sim 5.5$ Gpc.

\vskip .5cm

\noindent
\bigbox{II. ~Spherical shells}
\vskip .3cm

The above false-vacuum-black-hole configuration does not solve
Eq.\ (\ref{matter1}) at the horizon. Let us look for nontrivial
static solutions to Eqs.\ (\ref{matter1}) and (\ref{metric2}) which are
regular everywhere.  I show that such
solutions exist, if the  energy scale determined by the scalar
potential is sufficinetly small compared with
the Planck scale.\cite{Hosotani3,Dymnikova}  An explicit
example is given when $V(f_1)=\ep > 0$ and $V(f_2)=0$. The latter
condition
$V(f_2)=0$ can be relaxed.

\vskip -1.2cm 
\begin{figure}[h]
\centering \leavevmode 
\mbox{
\epsfxsize=16.cm 
\epsfbox{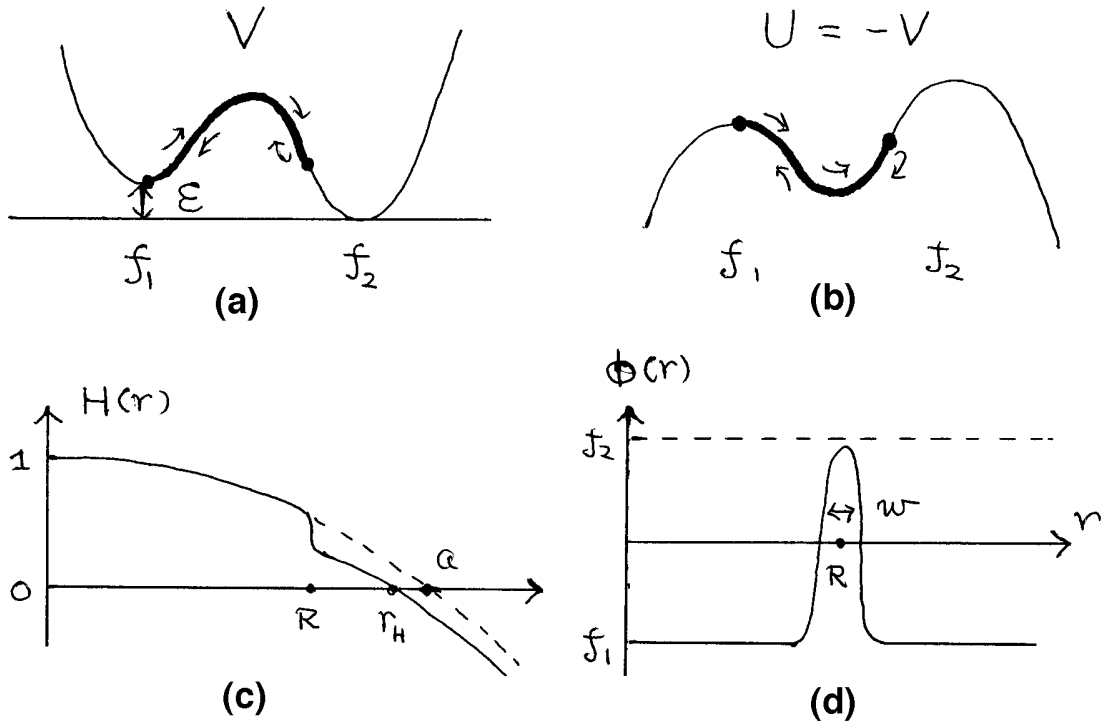}}
\caption{\baselineskip=10pt
Spherical shells in the de Sitter space.  (a) Scalar potential.
(b) Potential, $U[\phi]=- V[\phi]$, in the particle analogy.  
(c) $H(r)$ in the metric. 
(d) Schematic behavior of $\phi(r)$.  In the solutions $w/R \ll 1$.}
\label{fig1}
\end{figure}
\vskip .2cm

To see how such solutions become possible, we interpret Eq.\
(\ref{matter1}) as an equation for a particle with a coordinate
$\phi$ and time $r$.  Except for a factor $1/H$ a particle
is in a potential $U[\phi]=-V[\phi]$.  The coefficient $\Gamma_\eff(r)$
represents  time-dependent friction.  We are looking for
a solution  which starts at $\phi \sim f_1$, moves to $\sim f_2$, and
comes back to $f_1$ at $r=\infty$.  It is impossible in flat space,
as $\Gamma_\eff$ is positive definite so that the particle loses
an energy and cannot climb back to the original starting point.

In the presence of gravity the situation changes.  The non-vanishing
energy density can make $H$ decrease as $r$, and $\Gamma_\eff$ can
become negative.  The  lost energy of the particle during the 
initial rolling can be regained on the return path by  `negative
friction', or by thrust.  Indeed, this happens.

The schematic behavior of a solution is displayed in figure 1.
Take, as an example,  a quartic potential $V[\phi]$ with   
$|f_1|, |f_2| \sim 10^{-3} M_{\rm pl}$, and
$\ep/V_0 \sim 10^{-3}$ where $V_0$ is the  barrier hight.
In the solution $\phi$ starts at $\sim f_1$ and makes a transition
to $\sim f_2$ at $R \sim 10^7 \, l_{\rm pl}$.  The transition width
is about $w/R \sim 10^{-3}$.  [figure 1(d)]~  $H(r)$ in the metric
drops at $\sim R$. [figure 1(c)]

Why or how can such a configuration become possible?  First of all
I remark that there is no static regular solution which starts
at $\phi(0) \sim f_1$ and ends at $\phi(\infty) = f_2$.  In this
configuration there is a single transition from one minimum, $f_1$,
to the other, $f_2$, or one shell (wall).  Such a configuration is
unstable.  The shell collapses.

In the configuration depicted in figure 1, however, $\phi$ makes 
two transitions; $f_1 \go f_2 \go f_1$.  It defines two shells.
The energy density between the two shells is higher
than $\ep$, which gives rise to an attractive force between the shells.
There also is an intrinsic repulsive force between two domain
walls.\cite{Vilenkin}  Apparently these two forces balance each other
out to form static structure.  Yet, it is not clear how the
configuration is stabilized against shrinkage.   

The shell structure disappears as the energy scale of the scalar
potential becomes larger and approaches the Planck scale.  When
the energy scale, instead,  becomes smaller, a new type of  structure
with four or six shells  emerges.  

The global structure of the configuration is noteworthy.
In the metric of the  $R^1 \times S^3$ type,  a mirror of
shells appear in the other hemisphere of $S^3$.   As the universe
shrinks and expands, the intrinsic scale of the shells remain
constant.

Detailed analysis will be given in ref.\ \cite{Hosotani3}.

\vskip 1cm

\def\jnl#1#2#3#4{{#1}{\bf #2} (#4) #3}
\def\PRL{\em Phys.\ Rev.\ Lett. }
\def\PRB{{\em Phys.\ Rev.} B}
\def\PRD{{\em Phys.\ Rev.} D}
\def\PLB{{\em Phys.\ Lett.} B}

\leftline{\bf References}  

\renewenvironment{thebibliography}[1]
        {\begin{list}{[$\,$\arabic{enumi}$\,$]}  
        {\usecounter{enumi}\setlength{\parsep}{0pt}
         \setlength{\itemsep}{0pt}  \renewcommand{\baselinestretch}{1.2}
         \settowidth
        {\labelwidth}{#1 ~ ~}\sloppy}}{\end{list}}

\end{document}